\documentstyle[aps,epsfig]{revtex} 
\newcommand{\be}{\begin{equation}}
\newcommand{\ee}{\end{equation}\noindent}
\newcommand{\eei}{\end{equation}}
\newcommand{\bea}{\begin{eqnarray}}
\newcommand{\eea}{\end{eqnarray}\noindent}
\newcommand{\eeai}{\end{eqnarray}}

\newcommand{\hf} {{1\over2}}

\def\eq#1{(\ref{#1})}
\begin{document}
\draft
\title{On the renormalization of periodic potentials}
\author{I. N\'andori$^1$, J. Polonyi$^{2,3}$ and K. Sailer$^1$}
\address{$^1$ Department for Theoretical Physics, Kossuth Lajos University, 
Debrecen, Hungary}
\address{$^2$ Institute for Theoretical Physics, Louis Pasteur University,
Strasbourg, France}
\address{$^3$ Department of Atomic Physics, Lorand E\"otv\"os University, 
Budapest, Hungary}
\date{\today}
\maketitle
\begin{abstract}
The renormalization of the periodic potential is investigated in the 
framework of the Euclidean one-component scalar field theory by means of 
the differential RG approach. Some known results about the sine-Gordon
model are recovered in an extremely simple manner. There are two phases, 
an ordered one with asymptotical freedom and a disordered one where the 
model is non-renormalizable and trivial. The order parameter of the 
periodicity, the winding number, indicates spontaneous symmetry breaking 
in the ordered phase where the fundamental group symmetry is broken and 
the solitons acquire dynamical stability. It is argued that the periodicity 
and the convexity are so strong constraints on the effective potential that 
it always becomes flat. This flattening is reproduced by integrating out 
the RG equation.
\end{abstract}
\pacs{11.10.Hi, 11.10.Kk}

\section{Introduction}            
The dynamics generated by a periodic potential represents a challenge
in quantum field theory where the usual strategies to obtain the
solution are based on the Taylor expansion which violate the 
essential symmetry of the problem, the periodicity.
One expects two kinds of problems. A local one, the
perturbation expansion for the small fluctuations around a minima
of the potential should deal with infinitely many vertices in order
to preserve the periodicity. A global problem appears in 
the construction of the effective potential as an instability, 
a conflict between the periodicity and the convexity which can 
be resolved in a trivial manner only, being the constant the only 
function which is periodic and convex in the same time. Another 
global problem, made rather puzzling by the flatness of the 
effective potential, is to find an order parameter distinguishing 
the phase with periodicity from the
one where the periodicity is broken spontaneously.

The goal of this paper is to give a brief presentation of these
issues in the case of a two dimensional scalar model. The pertinent
features of the sine-Gordon model, the simplest realization of a 
periodicity, and the more detailed goals of this paper
are presented in section 2. Section 3 contains a 
very brief introduction into the Wegner-Houghton equation,
applied for the periodic potential in section 4. The linearized
renormalization group flow is given around the two-dimensional 
Gaussian fixed point and the periodic operators
of the potential are classified in section 5. The nonlinear
flow is discussed in section 6 by means of numerical integration of the
renormalization group equation. Section 7 contains a few remarks
about the signatures of the spontaneous breakdown of the 
fundamental group symmetry. Finally, section 8 is for the summary
of our findings.

\section{The sine-Gordon, X-Y and fermionic  models}
The simplest example for periodic potential is the two-dimensional
sine-Gordon model which is described by the Lagrangian
\begin{equation}\label{lagr}
L={1\over2}(\partial_\mu\phi)^2+V(\phi),
\end{equation}
with 
\begin{equation}\label{sgp}
V(\phi)=u\cos\beta\phi.
\end{equation}
The dynamics and the renormalization have been discussed by means of 
the straight perturbation expansion and the mappings between 
the sine-Gordon, the Thirring and the X-Y planar models
\cite{col}-\cite{agg}. The duality transformation
connecting the sine-Gordon and the X-Y models provides the renormalization
group flow in the complete coupling constant space \cite{kt,jkkn}.

The mappings between the different models are not exact. In fact,
the bosonization of the Thirring model \cite{col} and the 
Coulomb-gas representation of a periodic potential \cite{samu} are
obtained perturbatively. The duality is established up to irrelevant 
terms in the action \cite{jkkn}. But an exact equivalence exists between
the X-Y and the compactified sine-Gordon model which is 
obtained by expressing \eq{lagr} in terms of the compact variable
\cite{kerson}
\be
z(x)=e^{\beta\phi(x)}.
\ee
Such a parameterization makes the kinetic energy periodic,
\be
\label{clagr}
L={1\over2\beta^2}\partial_\mu z^*\partial_\mu z+{u\over2}(z+z^*)
\end{equation}
and introduces vortices in the dynamics.
The models \eq{lagr} and \eq{clagr} are equivalent in any order of the 
perturbation expansion in continuous space-time. The 
modification of the regulator amounts to the introduction of
non-renormalizable terms in the action which are irrelevant
in the UV scaling regime. Thus the equivalence is reached 
asymptotically only when the cutoff is removed. In lattice 
regularization the theory \eq{clagr} is in the same universality
class as the X-Y model which is described by the action
\begin{equation}
\label{actionXY}
S={1\over T}\sum_{<x,x'>}\cos(\theta_{x}-\theta_{x'})
+\sum_{x}h\cos(\theta_{x}),
\end{equation}
with $T=\beta^2$.
The renormalization of the X-Y model induces a non-trivial value
for the vortex fugacity $y$ which appears as an additional evolving
coupling constant in the compactified sine-Gordon model \cite{kerson}.
Therefore the complete coupling constant space contains three 
coupling constants, the external field $h$, the temperature 
$T$, and the vortex fugacity $y$. 

As it is well known, 
there are two phases in the $y-T$ plane connected by the 
Kosterlitz-Thouless transition \cite{kt}. In the low temperature, 
molecular phase $T<T_{KT}$ the vortices and anti-vortices form 
closely bound pairs while above the transition temperature $T>T_{KT}$, 
in the ionised phase they dissociate into a plasma. Due to the duality 
transformation the corresponding two phases appear in the $h-T$ 
plane as well as those of the dual electric Coulomb gas (DCG). 
In the $h-T$ plane the ionised phase of DCG is realised at 
low temperature, i.e. at weak coupling $\beta^2<\beta^2_c$ whereas the 
molecular phase of DCG is positioned at high temperature, 
i.e. at strong coupling $\beta^2>\beta^2_c$. Instead of
this duality transformation connecting the X-Y and the sine-Gordon
models we shall rely in this paper on the more direct equivalence
of \eq{clagr} and \eq{actionXY} as noted above \cite{kerson},
which identifies the weakly coupled (small $\beta$) and the
strongly coupled (large $\beta$) phase of the sine-Gordon model
with the molecular and the ionized phase of the X-Y model. The vortices
of \eq{clagr} and \eq{actionXY} correspond to each other in this scheme.

The transition between the phases can be characterized
either perturbatively or by means of the vortex dynamics. As far
as the perturbation expansion is concerned, the normal ordering is sufficient
to remove the UV divergences for weak couplings. Though the
divergence structure of the individual graphs is the same in either
phase the partial resummation of the perturbation expansion in $u$
produces new UV divergences for $\beta^2>8\pi$ \cite{col,bhn,sctr}. 
A further UV divergence was found
at $\beta^2=4\pi$ \cite{sctr}. The double expansion in $u$ and 
$\beta^2/8\pi-1$ \cite{agg} indicates no special singularities at
$\beta^2=4\pi$ and shows that the adjustment of $u$ and the introduction of
a wave function renormalization constant for the field $\phi$
in the strong coupling phase is sufficient to remove the UV divergences.

The inspection of the vortex dynamics allows us to follow the transition
line for larger values of $u$. One way the vortices arise is that
the vertices appearing in the perturbation expansion in (\ref{sgp}) form 
a gas of vortices described by the X-Y model \cite{samu}. 
Another way to identify the vortices is to use the 
equivalence of \eq{clagr} and \eq{actionXY} in lattice regularization
when the continuum limit is approached.
Both ways indicate that the weak and the strong coupling phases are
the molecular or the ionised phases, respectively from the point of view 
of the vortex gas and that the phases are separated by the 
Kosterlitz-Thouless transition. It is worthwhile noting that this
transition is rather peculiar because (i) its driving force, the 
vortex dynamics is generated by the UV modes rather than the 
IR ones as in the case of spontaneous symmetry breaking, and (ii)
it is a higher than second order since the correlation length is
infinite in the molecular phase and diverges faster than  
any power of the reduced temperature (exponentially), therefore 
one can not introduce the critical exponent $\nu$ in the usual manner.

The sine-Gordon model possesses a topological current, 
\begin{equation}
j_\mu(x)={\beta\over2\pi}\epsilon_{\mu \nu}\partial_\nu\phi(x)
\end{equation}
which is conserved
in the semi-classical expansion, when the path integral is saturated by 
field configurations with analytic space-time dependence. The flux
defined by $j_\mu$ is the vortex number, called the vorticity and 
the soliton number in the same time. In this manner the world 
lines of the sine-Gordon solitons end at the X-Y model vortices, 
making the soliton unstable and the topological current anomalous 
in the ionised phase where field configurations with singular
space-time dependence survive the removal of the cutoff \cite{kerson}. 
This destroys or at least modifies the bosonization transformation in 
the high temperature, ionised phase of the vortex gas. In fact, 
the non-conservation of the
topological current requires fermion number non-conserving terms
in the fermionic representation, a fundamental violation of the
rules inferred from the weak coupling expansion.

The relation between the sine-Gordon and the X-Y model is summarized 
in table \ref{XY}.

\section{Differential RG approach in momentum space}
The challenge in developing an RG method for the sine-Gordon model
is that it should follow the mixing of all the operators which become
relevant in either phase. Since there are infinitely many
relevant operators in two dimensions one needs the functional form
of the evolution equations for the blocked Wilsonian action \cite{wh}-\cite{int}.
We shall use the leading order gradient expansion in the Wegner-Houghton equation
\cite{wh} to study the renormalization group flow of a generalized model with
an arbitrary periodic potential. Such a drastic truncation leaves some
doubts due to the supposed role of the wavefunction renormalization constant
in the ionised phase. But the higher order contribution of the
gradient expansion can only be treated consistently by the use of the effective
action instead of the bare one. We shall find that the instability arising from
the periodicity of the effective potential makes the Legendre transformation
highly nontrivial and prevents us to use the effective
action in the infrared regime of the molecular phase. 

The differential RG transformations are 
realised by integrating out the high-frequency Fourier components of the field 
variable, in infinitesimal steps in momentum space successively from the 
UV cut-off $k$ to $k-\delta k$,
\begin{equation} 
\label{path}
e^{-S_{k-\delta k}[\phi]} = \int D[\phi'] e^{-S_{k}[\phi+\phi']}, 
\end{equation}
where $S_k[\phi]$ stands for the blocked action with cutoff $k$
and the field variables $\phi$ and $\phi'$ contain
Fourier components with momenta $p<k-\delta k$, 
and $k-\delta k<p<k$, respectively. In every 
infinitesimal step, the path integration in Eq.(\ref{path}) is evaluated by the 
help of the saddle point approximation. If the saddle point is at $\phi'=0$, 
one can find an integro-differential equation for the blocked action, called the 
Wegner-Houghton equation \cite{wh}. In the local potential approximation
one uses the leading order expression for the action in the gradient
expansion,
\begin{equation}
\label{lopo}
S_k=\int d^dx\left[{1\over2}(\partial_\mu\phi(x))^2+V_k(\phi(x))\right],
\end{equation}
and the Wegner-Houghton equation reduces to a differential equation 
for the scale dependent potential $V_k(\phi)$ \cite{senben}
\begin{equation} 
\label{WH}
k\partial_k V_k(\phi)=-k^d\alpha\ln\left(k^2+\partial^2_{\phi}V_k(\phi)
\right),
\end{equation}
with $\alpha={1\over2}\Omega_d(2\pi)^{-d}$, and the solid angle $\Omega_d$ in 
dimension $d$.

Notice that the argument of the logarithm in \eq{WH} must be
non-negative for the expansion made around a stable saddle point.
If the argument becomes negative at a critical value 
$k_{cr}>0$, given by 
$k_{cr}^2 = -\partial^2_{\phi} V_{k_{cr}}(\phi)$  then the Wegner-Houghton equation 
looses its validity for $k<k_{cr}$ and the saddle point becomes 
non-zero and the tree-level blocking relation must be used. 
We simplify the saddle point structure of the blocking \eq{path}
by retaining the plane waves only and we find the evolution equation
\begin{equation} 
\label{tree}
V_{k-\delta k}(\phi) = \min_{\rho} \left[k^2 \rho^2 + {1\over 2} \int_{-1}^{1} du
V_{k}(\phi + 2\rho \cos(\pi u)) \right],   
\end{equation}
where $\rho$ is the amplitude of the plane wave \cite{tree}.

\section{Periodicity of the potential}
The symmetry of the action under the transformation
\be\label{pertr}
\phi(x)\to\phi(x)+\Delta
\ee
is to be preserved by the blocking and the potential $V_k(\Phi)$
must be periodic with period length $\Delta$. It is actually obvious
that the blocking, the transformation 
\begin{equation} 
\label{WH per}
kV_{k-\delta k}(\phi)= kV_k(\phi)+\left[k^d\alpha
\ln \left(k^2+\partial^2_{\phi} V_k(\phi)\right)\right]\delta k
\end{equation}
preserves that periodicity of the potential. The impact of the 
periodicity of the potential on the dynamics can be understood 
by recalling that $V_k(\Phi)$ tends to the effective potential,
$V_{eff}(\Phi)$ in the IR limit, $k\to0$. This is because both
potential give the action density within the functional space
$\Phi=\langle\phi(x)\rangle$. The Legendre transform imposes
the condition of the convexity on the effective potential
\cite{conv} and thereby on $V_{k=0}(\Phi)$. 
Since the only periodic and convex function is the
constant there is no way to have nontrivial effective potential
when the transformation \eq{pertr} is a 
formal symmetry of the action. Notice that this statement holds
for any dimensions. The detailed RG study of the sine-Gordon model
is pursued here to check such a general conjecture in a simpler case.

One could object by pointing out that it is enough to implement the
horizontal Maxwell cut of the effective potential between the 
inflexion points $V''_{eff}(\Phi)=0$, where the function is concave.
But the problem left by this construction is just at the inflection
point where the potential obtained in this manner has singular higher derivatives.
This is unacceptable, as one can see by placing the system in a
large but finite box. On one hand, the effective potential should
be close enough to the one obtained in the thermodynamical limit,
i.e. the Maxwell cut should be present up to finite size corrections.
On the other hand, the effective potential should be regular in the
absence of IR divergences. It is easy to see that the singularity
at the inflexion point gives rise to smooth but concave region
for finite systems after being rounded off and the Maxwell cut must be 
extended further. In the case of the $\phi^4$ model where the symmetry
which is broken spontaneously is $\phi\to-\phi$ such an extension
of the Maxwell cut leads to the degeneracy of the effective potential
between the minima \cite{tree}. When the symmetry \eq{pertr} is broken then
infinitely many minima are expected in the effective potential. The
only Maxwell cut which is stable in the thermodynamical limit is thus
the one which renders $V_{eff}(\Phi)$ constant.

In order to allow a convenient truncation of the potential for the numerical
solution of the evolution equation and to preserve the periodicity
we write the $V_k(\phi)$ as a Fourier series, 
\begin{equation}
V_k(\phi)=\sum_{n=0}^{\infty} u_n(k) \cos\left(n\beta\phi\right).
\end{equation}
For the sake of simplicity we consider only potentials with  
Z(2) symmetry, $V_k(\phi)= V_k(-\phi)$. The whole scale dependence
occurs in the Fourier amplitudes  $u_n(k)$, the `coupling constants' of the 
scale dependent potential.
In the case of a non-trivial saddle point Eq.(\ref{tree}) can be rewritten as
\begin{equation} 
\label{tree per}
V_{k-\delta k}(\phi) = \min_{\rho} \left[k^2 \rho^2 + \sum_{n=1}^{\infty} 
u_n(k) \cos(n \beta\phi) J_{0}(2 n \beta  \rho)\right],   
\end{equation}
where $J_{0}$ stands for the Bessel function and $\beta={2\pi/\Delta}$. 
Let $\rho_k(\phi)$ denote
the position of the minimum of the bracket on the r.h.s. of Eq.(\ref{tree per}). 
Then $\rho_{k}(\phi)$ is periodic $\rho_{k}(\phi+\Delta)=\rho_{k}(\phi)$, 
since one has to minimise the same expression of $\rho$ for $\phi$ and 
$\phi+\Delta$. Note that Eq. \eq{tree per} preserves the periodicity.

The retaining of the higher order contributions in the gradient expansion, 
among them the leading order being the inclusion of a wavefunction 
renormalization constant $Z_k(\phi(x))$ into the
kinetic energy of the action (\ref{lopo}) changes the situation.
The action keeps its period length in the bare field under any circumstance.
But the period length in terms of the renormalized field 
$\phi_{R,k}(x)=Z_k^{1\over2}(\phi_0)\phi(x)$, where $\phi_0$ minimizes the
potential $V_k(\phi)$, is $\Delta Z_k^{1\over2}(\phi_0)$. Such sub-leading
contributions of the gradient expansion are neglected in the
present work. We believe that their contribution will not change 
our results qualitatively.

It is easier to use the derivative of Eq.(\ref{WH}) with respect to $\phi$ 
rather than the original equation itself. For dimension 
$d$ the general form of the evolution equation reads as follows,
\begin{equation}
\label{general}
\alpha \beta^{2} k^{d-2} n^2 v_n(k) = (d+ k\partial_k) v_n(k) - 
{1\over 2} \beta k^{d-2} \sum_{p=1}^{N} A_{n p}(k) (d+ k\partial_k)v_p(k), 
\end{equation}
where $N$ is the truncation in the Fourier series, $v_n(k) = n\beta u_n(k)$  and 
\begin{eqnarray}
A_{n p}(k)&=& (n-p)v_{n-p} \Theta(n\geq p) \\ \nonumber
          &+& (p-n)v_{p-n} \Theta(p\geq n) \\ \nonumber
          &-& (n+p)v_{p+n} \Theta(N\geq n+p),
\end{eqnarray}
with $\Theta(n\geq n') = \{1 $ if $ n\geq n', 0 $ if $ n<n' \}$.

Reaching the critical value $k_{cr}$ one has to change automatically from the 
system of equations (\ref{general}) to Eq.(\ref{tree per}).
In every step $\delta k$ in the momentum space, the potential at the scale
$k-\delta k$ is then found by minimising the expression on the r.h.s. of 
Eq.(\ref{tree per}). After the minimisation, the potential $V_{k-\delta k}(\phi)$
is expanded in Fourier-series to define the new Fourier amplitudes at the scale
$k-\delta k$. One repeats this algorithm step by step till $k=0$.

\section{Linearized solution}
According to the power counting, theories with polynomial interactions are 
super-re\-nor\-ma\-li\-zab\-le in dimension $d=2$. Furthermore, the 
super-renormalizable interactions correspond to relevant operators in 
the UV scaling regime. How can we have new UV divergences in the
ionised phase when the set of the renormalizable operators is fixed?
The source of the complication is that in the usual perturbative proof of the 
renormalizability each monomial vertex is treated
independently. This strategy is sufficient for polynomial interactions but is not 
necessarily applicable for periodic potentials where the symmetry is 
destroyed by any truncation of the Taylor expansion.

The treatment of an infinite series of operators instead of a
single monomial may cause complications, an impression of
having renormalized a manifestly non-renormalizable model. In fact, 
we may find an infinite series of irrelevant operators in a
renormalizable model, showing the possibility of the removal
of the cutoff in the presence of non-renormalizable operators.
An obvious example is when a regulator, represented as an interaction 
vertex, yields irrelevant operators. We find this situation by
introducing the finite difference operator appearing in the 
lattice regularized theories in the continuum. The finite
difference operator generates an infinite power series of the 
gradient whose monomials are non-renormalizable. 
The basic question in the removal of the cutoff is whether the series
of the irrelevant operators is chosen in such a manner
that the divergences can be removed by the fine tuning of a finite 
number of 
parameters in the action. The infinite series of irrelevant operators
is usually required by some symmetry of the theory, such as the
periodicity in momentum space on the lattice  \cite{reiszs},
the global $O(N)$ symmetry of the nonlinear sigma model \cite{nlsig}
or gauge theories on the lattice \cite{reiszg}. The symmetry
imposes such constraints on the radiative corrections that the
divergences can in fact be removed within the given functional 
family of the action and the apparently non-renormalizable model
becomes renormalizable.

One touches upon here a fundamental difference between the
ways renormalization group is used in statistical mechanics and particle
physics. In statistical mechanics the UV cutoff is physical and we may not
ignore the effects taking place at that scale. In particle physics we insist
that the cutoff is sufficiently far from the scale of the phenomenon we are
interested even in the effective theories. The gain coming from this
constraining of the set of observables is that universality 
arguments apply and it is enough to
consider renormalizable models\footnote{The universality might need certain
generalization in case of several scaling regimes or instabilities
\cite{tree}, \cite{glob}.}. But we may need irrelevant operators 
in statistical mechanics which lead to complicated non-renormalizable
models. The error caused by the omission of the irrelevant pieces in the
particle physical applications is negligible. We can turn this 
insensitivity into a freedom: The suppression mechanism
responsible of this simplification gives us the possibility to include
irrelevant pieces in the theories without specific fine tuning of their
coupling constants in particle physics so long as the UV divergences can be
removed. 

In the case of the periodic potential we have the opposite
effect, a restriction on the renormalizability due to the presence of
infinitely many vertices in the model. This is because the new UV divergences
of the ionised phase arise from the summation of 
infinitely many, individually finite graphs \cite{bhn}, 
\cite{sctr} ,\cite{agg}. We show that this highly nontrivial effect can be
reproduced in a very simple manner in the framework of the functional form of
the renormalization group method. We shall use the Fourier amplitudes $u_n(k)$ as
coupling constants and we linearize the $d=2$ dimensional
renormalization group flow (\ref{general}) around the fixed point 
$u_n=0$ by assuming $\vert\partial^2_\phi V_k(\phi)\vert\ll k^2$.
Notice that the UV Gaussian fixed point is well-defined in any 
dimensions because modes with non-vanishing momentum
are considered in the blocking. The singularity building up at or below 
the lower critical dimension can be found in the flow generated by such a blocking
procedure as an instability and as an inconsistency of the massless
IR fixed point without influencing the UV scaling laws.

It is more natural to express the flow in terms of the dimensionless coupling 
constants, $\tilde u(k)=k^{-d} u(k)$. The solution of the linearized
renormalization group equation satisfying the initial conditions 
$\tilde u_{n}(k=\Lambda)=\tilde u_{n\Lambda}$ is
\begin{equation}
\label{lind2}
{\tilde u_n(k)}={\tilde u_{n\Lambda}}
\left({k\over\Lambda}\right)^{\left(\alpha {\beta^2} n^2 -2\right)}.
\end{equation}
Thus, the coupling constants $\tilde u_n(k)$ are relevant, marginal, or 
irrelevant for $\beta^2 < 8\pi/ n^2$, $\beta^2 = 8\pi/ n^2$, 
or $\beta^2> 8\pi/ n^2$, respectively.

For $\beta^2 > 8\pi$ all dimensionless coupling constants are irrelevant
and (\ref{lind2}) is consistent and keeps the trivial saddle
point of the blocking stable. This result indicates the inaccessibility
of the Gaussian fixed point in the UV limit, the non-renormalizability 
of the model. The infrared fixed point
is a trivial, non-interacting massless theory.

It is instructive to compare this result with the 
$O(u^3)$ prediction of the double expansion in $u$ and 
$\beta^2/8\pi-1$ of the sine-Gordon model which shows the 
possibility of absorbing the UV divergences into $u$ and a
wavefuntion renormalization constant for the field \cite{agg}. 
The doubt that the higher order contributions of the perturbation 
expansion may render the theory non-renormalizable is resolved
in ref. \cite{agg} by relying on the bosonization method.
Unfortunately there are problems with the bosonization in the ionised 
phase which is not surprising since this mapping is based either on
the saddle point or the weak coupling expansion. The problem is the anomaly of
the topological current which requires the presence of fermionic operators
with odd power in the fermion field \cite{kerson}. We believe that
the theory is non-renormalizable in this phase but the real proof
requires further efforts.

The potential becomes flat under RG transformation in the limit $k\to0$ 
and the flat effective potential suggests the presence of 
a massless particle which is in contradiction with the dimensionality 
of the system \cite{mwc}. The resolution of this apparent contradiction 
is that according to \eq{lind2} the non-Gaussian vertices of 
the model tend to zero sufficiently 
fast in the infrared limit to suppress the IR divergences
of the perturbation expansion
when each mode in the loop integral is coupled to the
effective coupling constants at the appropriate scale.

The critical point at $\beta^2=8\pi$ is a well-known result for the 
sine-Gordon model. Furthermore it was known that the higher
harmonics, corresponding to the vortices with higher vorticity
are irrelevant around the critical point \cite{agg}. What is interesting in 
our solution is that one sees a change in the scaling laws for the
higher harmonics at a finite distance away from this critical point.
In fact, the $n$-th harmonic is found to be relevant for 
$\beta^2\le8\pi/n^2$. The apparency of a new relevant 
operator implies not only a new renormalization condition but
the possibility of new ultraviolet divergences. Though we 
find no evidence for the singularity at $\beta^2=4\pi$ 
\cite{sctr}, a series of new singularities is expected
at $\beta^2=8\pi/n^2$, due to the vortices with higher 
vorticity.

\section{Numerical solution}
For $\beta^2<8\pi$ the first few Fourier-amplitudes are relevant, that is 
they increase for decreasing value of $k$ and consequently the linearisation 
ceases to be reliable. The solution can only be found numerically in this case.
Monitoring the coupling constants numerically we compare
the solution of Eq.(\ref{general}) with the result obtained from the 
analogous relation for the polynomial potential \cite{tree}, 
\begin{eqnarray}
V_k(\phi)=\sum{1\over n!}g_n\phi^n.
\end{eqnarray}
The initial conditions for the polynomial potential were chosen $g_2=-0.001$, 
$g_4=0.01$ and $g_n=0$, if $n>4$. Then the saddle point remains trivial
for any $k$ and the singularity of Eq.(\ref{WH}) at 
$k^2_{cr}(\phi)=-\partial^2_{\phi} V_k(\phi)$ is avoided.
In order to compare the periodic and the polynomial cases we choose 
the initial conditions for the Fourier-amplitudes $u_n(k=\Lambda)$ for 
$\beta^2 = 0.1 \beta_c^2$ such, that after Taylor-expansion the initial 
conditions for the polynomial case are recovered. Therefore, the initial 
conditions for various truncations $N$ of the Fourier-series are different. 
For the increasing values of $N$ the coupling constants $g_n(k)$ 
determined from the periodic potential by the Taylor-expansion approach
the running coupling constants of the polynomial potential in 
the UV regime. This is understandable since the quantum field $\phi$ 
does not feel the global properties of the potential, i.e. the periodicity
due to its small fluctuations in the UV regime. Just the opposite holds in the IR 
regime, where the field fluctuations become larger and they make
the global features of the potential manifest. Therefore, it is
expected that the solutions for the periodic and the polynomial potentials 
become different in the IR regime.   

\begin{figure}
  \vspace{-2cm}
  \epsfig{file=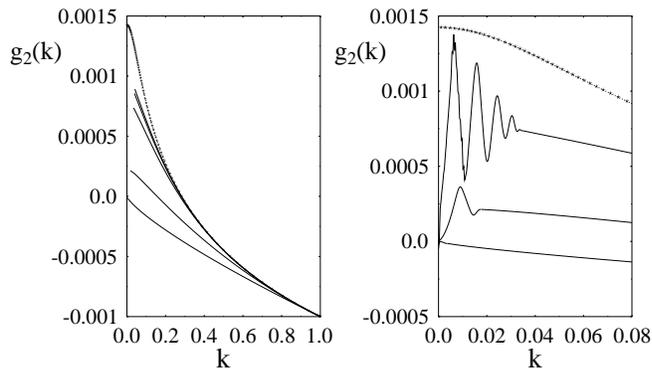, width=8.6cm}
  \vspace{-2cm}
  \caption{Comparison of the scaling of the dimensionful coupling constant 
    $g_2$ obtained for the polynomial (dashed-line with stars) 
    and for the periodic potential for various values of the truncation 
    $N=2,3,10,20,30$ in the Fourier series. The figures on the left (right) 
    show the scaling of the coupling above (below) $k_{cr}$. 
    Below $k_{cr}$ we show the scaling of $g_2$ for the cases $N=2,3,10$. 
    The increasing order of the full-line curves corresponds to increasing $N$.}
  \label{g2}
\end{figure}

\begin{figure}
  \vspace{-2cm}
  \psfig{file=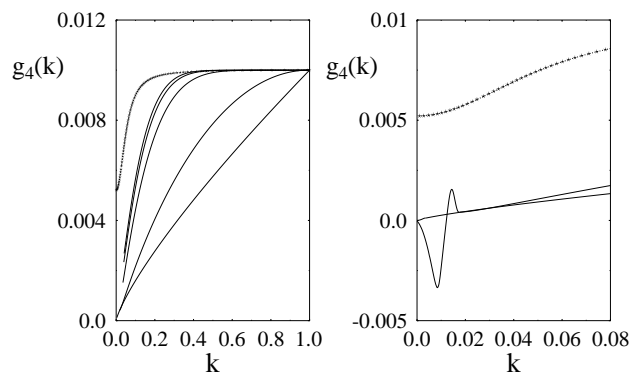, width=8.6cm}
  \vspace{-2cm}
  \caption{The scaling of the dimensionful coupling constant
    $g_4$ in the same cases as for $g_2$ in Fig.\ref{g2}. The figures on the 
    left (right) show the scaling of the coupling above (below) $k_{cr}$.}
  \label{g4}
\end{figure}

We integrated numerically Eq.(\ref{general}) starting from the UV cutoff 
$\Lambda=1$ down to the critical value 
$k^2_{cr}(\phi)=-\partial^2_{\phi}V_{k_{cr}}(\phi)$ by using the fourth-order 
Runge-Kutta method and $\delta k=10^{-p} k$ with $p=3$ or $p=4$. 
There were no changes in the numerical results by increasing $p$
further. In Fig.\ref{g2} and Fig.\ref{g4} we show the scaling of the 
dimensionful coupling constants $g_2$ and $g_4$ for different truncations 
$N$. Increasing the value of $N$ the differences in the results obtained 
for the periodic potential decrease.  

There are relevant coupling constants for $\beta^2<8\pi$ 
which become large enough to destabilize the trivial saddle point
of the blocking and we reach a non-vanishing critical value $k_{cr}$
where the saddle point becomes non-trivial and the tree-level blocking 
Eq.(\ref{tree}) must be used. By following the solution of this 
equation all dimensionful Fourier-amplitudes are found to approach 
zero as $k\to0$. The typical behavior is depicted in Fig.\ref{g2} 
and Fig.\ref{g4}.
 
\begin{figure}
  \epsfig{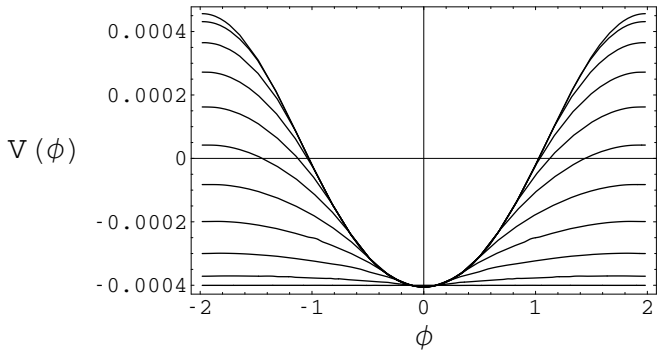}
  \vspace{0.5cm}
  \caption{Flattening of the periodic potential below $k_{cr}$. The decreasing
    order of the curves corresponds to decreasing values of the scale $k$
    with the step $\delta k= k_{cr}/10$.}
  \label{pot}
\end{figure}

There is a remarkable difference in the behavior of the theory with  
a periodic potential and that with the corresponding polynomial potential
\cite{lowd}. 
Namely, that all the dimensionful coupling constants $g_n(k)$ obtained for 
the periodic potential tend to zero in contrary to those of the
polynomial potential which remain finite as $k \to 0$.
In Fig.\ref{pot} we show this flattening starting from the value $k_{cr}$.

\begin{figure}
  \epsfig{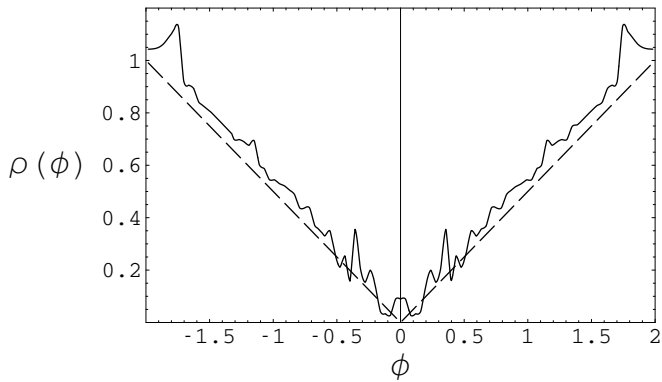}
  \vspace{0.5cm}
  \caption{Comparison of the function $\rho_{k=0}(\phi)$  obtained by numerical 
    integration of Eq.(\ref{tree per}) with the analytic expression 
    $\rho_{k=0}^{per}(\phi) = {1\over 2} \vert\phi\vert$ (dashed-line).}
  \label{rho}
\end{figure}

Integrating numerically Eq.(\ref{tree per}) we have shown that 
under the successive infinitesimal RG transformations the periodic 
potential becomes a constant potential up to the accuracy 
$10^{-5}$. We compared our results with the analytic formula for the 
saddle point amplitude $\rho_{k}(\phi)$ obtained for the polynomial 
potential \cite{tree}:
\begin{equation}
\label{rho-poly}
\rho_{k}(\phi)={1\over2}\left(\phi_{vac}(k)-\vert\phi\vert\right).
\end{equation}
For periodic potentials $V(\Phi+\Delta)=V(\Phi)$, the amplitude 
$\rho_{k}(\phi)$ should be periodic in the field variable with the 
length of period $\Delta$ for any scale $k$ (see Eq.(\ref{tree per})).
Thus, for periodic potentials, an expression similar to Eq.(\ref{rho-poly}) 
is valid in the period $\phi \in [-\Delta/2,\Delta/2]$ and then the same 
pattern of the function $\rho_{k}(\phi)$ is repeated in all other 
periods. In the particular case investigated by us the minus sign in 
expression (\ref{rho-poly}) should be changed to a plus sign
and the field independent term is $\phi_{vac}^{per}(k_{cr})=-\Delta/2$ 
and $\phi_{vac}^{per}(0)=0$. Therefore we can compare the 
result $\rho_{k=0}^{per}(\phi)=\hf\vert\phi\vert$ 
with that obtained by numerical integration of Eq. \eq{tree per}, 
see Fig.\ref{rho}. We have established that with the increasing number of 
Fourier-modes taken into account the computed curves get closer to the 
dashed line defined by $\rho_{k=0}^{per}(\phi) = {1\over 2} \vert\phi\vert$.

\section{Spontaneous symmetry breaking}
We found two different phases of the model, $\beta^2<8\pi$ and
$\beta^2>8\pi$, for small $u$ with different scaling laws.
The effective potential flattens out in either case. 
How can we reconcile this result with the usual perturbation 
expansion of the sine-Gordon model \cite{col} where one expects
small fluctuations around one of the minima of the periodic potential
for $\beta^2<8\pi$? What does happen with the symmetry $\phi\to\phi+2\pi$? 
The flattening of the effective potential makes the 
naive order parameter to detect the realization of the periodicity, 
$\int dx\phi(x)$, useless. What is the corresponding order parameter?

The common difficulty behind these questions is that the 
physical configuration space is actually multiply
connected. In fact, the sine-Gordon Hamiltonian describes
a family of coupled
pendulums and the configurations $\phi(x)$ and $\phi(x)+\Delta$
are physically indistinguishable. We may use either the 
multiply connected description in terms of the complex
variable $e^{2\pi i\phi(x)/\Delta}\in U(1)$ or the
unconstrained $\phi(x)\in{\cal R}$ variable on the covering space
for the description of the system, and the transformation \eq{pertr}
belongs to the fundamental group $\pi_1(U(1))=Z$.

The special difficulty in detecting the spontaneous
breakdown of the fundamental group symmetry is the degeneracy
of the vacuum energy density, the only resolution of the 
conflict between the periodicity and the convexity of 
the effective potential in either phase.
The flattening of the effective action in the region of the
Maxwell-cut indicates the
weakness of the restoring force acting on the fluctuations
around the equilibrium position and raises the possibility
of the collapse of the topological stability of the solitons.
This, in turn, suggests the winding number as an order parameter
to distinguish between the phases with the usual and the unusual
realization of the fundamental group symmetry. This circumstance
reflects a further similarity with gauge models at finite temperature
\cite{deconf}. 

To understand these questions better we start by distinguishing
the local potential of the effective theory $V_k(\phi)$ with a low cutoff 
$k$ from the effective potential, $V_{eff}(\phi)=V_{k=0}(\phi)$. 
In theories with infrared stable
dynamics the limit $k\to0$ is safe and this difference is
negligible. But the infrared instability inducing the mixed phase
and the Maxwell cut in the effective potential makes the limit $k\to0$
more involved even for theories with massive particles only 
\cite{tree}. By turning this complication into an advantage,
we suggest to consider the infrared instabilities as 
the signature of the spontaneous symmetry breaking. 

Let us consider the dynamics of the modes with
momentum $p<k\not=0$. The local potential $V_k(\phi)$ of the corresponding
effective theory is non-trivial and periodic and the perturbation expansion 
might be justified around one of the minima for $k>0$. 
The real question is the relative speed the different
coupling constants approach 0 as $k\to0$, and whether the 
saddle point and the perturbation expansion remain consistent 
in this limit. As mentioned above, we adopt the infrared instability, 
i.e. the stability of the trivial saddle point of the blocking as 
a signature of the absence of the change of the symmetry pattern of the model. 

The saddle point of the blocking remains trivial in the ionic
phase and we expect no spontaneous symmetry breaking there.
The coupling constants decrease as $k\to0$, the theory
becomes trivial as it happens with other non-renormalizable models.
The infinitesimal fluctuations gradually fill 
up the valleys of the local potential and the naive
order parameter for the periodicity, $\int dx\phi(x)$, decouples.

The large amplitude, tree level fluctuations which lie 
beyond the realm of the perturbation expansion
but are picked up by the tree-level renormalization start to
fill up the valleys of the potential in the molecular
phase already at a finite scale, $k<k_{cr}$. We take this
instability as the indication of spontaneous symmetry
breaking. This seems to be in agreement with the perturbative 
approach which is meaningful only when the potential has a
non-trivial structure. 

In order to identify the order parameter for the periodicity we 
elucidate the topological differences between the two phases.
Let us introduce periodic boundary conditions in the 
Euclidean space-time and consider the following quantities.
One of them is the winding number,
\begin{eqnarray}
Q(x^0)=\int dx^1j_0(x^0,x^1),
\end{eqnarray}
the space integral of the topological current density at a given time, $x^0$.
We shall argue that this is the order parameter for the fundamental
group symmetry. One can construct another, independent topological
invariant by exchanging the time and the space axes. But this does not
modify the discussion what follows. The other quantities are the vorticities of the 
space-time regions before and after the time $x^0$,
\begin{eqnarray}
V_\pm=\int_{\pm(x^0-y^0)>0}d^2y\partial_\mu j_\mu(y).
\end{eqnarray}
The flux of the topological current agrees with the 
vorticity of the enclosed vortices $Q=V_-=-V_+$ \cite{kerson}.

In the molecular phase the  distance between the
vortex-anti vortex pairs is shrinking with the increasing UV cutoff
and the quantum fluctuations cannot change the value of
$Q$. This makes the path integration consistent when constrained
within a homotopy class, characterized by a fixed value of $Q$.
In fact, the consistency of the path integration is the
requirement that the path integral 
as the function of the end point of the trajectories should
satisfy the (functional) Schr\"odinger equation. The Schr\"odinger
equation can be derived for the path integral by performing
infinitesimal variations on the trajectories at the 
final time. Thus the path integral is consistent if the
functional space over which it is evaluated is closed
with respect to infinitesimal deformation of the trajectories.
Since the discontinuous field configurations
are suppressed in the path integral the quantization process
is well defined in a given homotopy class \cite{deconf}. 
Note that the consistent
constraining the functional integration into a given
homotopy class removes the fundamental group symmetry.
This is obvious in the semi-classical quantization of a soliton
where the dynamical stability of the whole construction comes
from the spontaneous breakdown, the suppression of the winding number
changing processes. Thus the condition for the stability of the solitons, 
the sufficient smoothness of the configurations dominating 
the path integral, is in the same time the signature of 
the breakdown of the fundamental group symmetry.

There are vortex-anti vortex pairs with cutoff
independent separation in the ionised phase and $Q$
fluctuates in an uncontrollable manner. The path integration
cannot be constrained into a given homotopy sector,
the fundamental group symmetry is realized but the 
semi-classical structure based on smoothness is destroyed,
the solitons become unstable. In the same time the
bosonization relations are either lost or fundamentally
modified because the soliton
(fermion number) non-conserving processes require the 
introduction of operators with odd powers of the
fermion field in the action.

The susceptibility of the topological charge, 
\begin{eqnarray}
\chi=\langle Q^2\rangle-\langle Q\rangle^2
\end{eqnarray}
may serve as a disorder parameter to distinguish the
different realizations of the periodicity. In fact, the 
vortex fugacity tends to zero or stays finite in the molecular 
phase or the ionized phase when the cutoff is removed according 
to the Kosterlitz-Thouless RG flow. Since the topological 
susceptibility is vanishing whenever the fugacity is zero, 
the former may serve as an order parameter.

\section{Summary}
A simple two dimensional scalar model with periodic potential was
investigated in this paper. The first question considered was
the renormalization of the potential. The study of this problem
requires the handling of infinitely many operators which 
was achieved by the Wegner-Houghton equation. We found a
disordered phase where the model is non-renormalizable and trivial.
In another, ordered phase the relevant operators
compatible with the symmetry were identified.

The periodicity and the convexity impose triviality on the
effective potential, a phenomenon verified in detail in both
phases. The coupling constants approach zero regularly
in the disordered phase as the theory becomes trivial,
leading to the flattening of the effective potential.
Instabilities were found in the ordered phase where the
effective potential is flattened out by the Maxwell
construction only, indicating a nontrivial dynamics
behind the trivial final result.

It was pointed out that the configuration space of the
model is multiply connected and that the winding number 
is suggested as a non-local order parameter distinguishing
the explicit realization and the spontaneous breakdown
of the fundamental group, the shift of the field variable
by the period length of the potential. The relation between
the applicability of the semi-classical arguments and the
spontaneous breakdown of the fundamental group symmetry is
shown in the context of our model.

We have started to extend this work to include the 
wavefunction renormalization constant for the field,
and to investigate the phase structure in higher dimensions.
The scalar model with periodic potential can be considered as a
non-linear $O(2)$ model with a symmetry breaking term, and one expects
similarities with non-Abelian gauge theories due to the compact
nature of the dynamical variable. In fact, the molecular phase is
asymptotically free and the vacuum is filled up with vortices in its 
large distance structure.
Furthermore, the spontaneous breakdown of the fundamental group symmetry
is reminiscent of the breakdown of the center symmetry in finite temperature
gauge theories. Thus, our results offer interesting lessons to be learned 
in constructing the quark confinement mechanism.

We ignored the wavefunction renormalization constant in this
work. In order to go beyond this approximation and to take into
account some higher order contributions in the gradient expansion
one has to turn to the evolution equations for the effective
action, instead of the bare action \cite{wett}, \cite{morr}, \cite{int}. 
But the problem is just that the effective action, appearing in the renormalization
group treatment, hides a large part of the dynamics due to
the Maxwell cut. This is because the effective action, obtained
by a smooth cutoff is after all the Legendre transform of the
logarithm of a path integral which does contain the IR modes.
These modes, though they appear with small amplitude, 
generate the Maxwell cut. It is not clear to us how to improve the
renormalization group method to make it applicable for
models with spontaneous symmetry breaking or with 
compact variables (e.g. gauge models with gauge fixing which are
based on compact gauge group) beyond the local potential
approximation.

\acknowledgments One of the authors (I.N.) thanks J. Alexandre, 
V. Branchina, G. Plunien for the useful discussions . This work is 
supported by the NATO grant PST.CLG.975722, the DAAD-M\H OB 
project $N^o-27/1999$, and the grants OTKA T023844/97, T29927/98.
One of the authors (K.S.) thanks the Alexander von Humboldt Foundation
for the follow-up fellowship and W. Greiner for his kind hospitality 
allowing to achieve a good progress in this work.

%
%

\newpage

\newpage 

\mediumtext
\begin{table}[t]
\begin{tabular}{|c|c|}
\hline
XY-model with external field & compactified sine-Gordon model\\
\hline\hline 
external field $\>\>\> h$
&Fourier amplitude $\>\>\> u=h/\beta^2 \>\>\>\>$\\
\hline
temperature $\>\> T$ 
&coupling constant $\>\>\> \beta^2 \>\>\>\>\>$\\
\hline
molecular phase of the vortex gas  $\>\>\> T<T_{c} \>\>\>\>$
&weak coupling phase $\>\>\> \beta^2<\beta^2_{c} \>\>\>\>\>\>\>\>\>$\\
\hline
ionised phase of the vortex gas  $\>\>\> T>T_{c} \>\>\>\>$
&strong coupling phase  $\>\>\> \beta^2>\beta^2_{c} \>\>\>\>\>\>\>\>\>$\\
\hline
vortex, anti-vortex
&soliton creation and annihilation\\
\hline
\end{tabular}
\caption{Comparison of the XY and the sine-Gordon model}
\label{XY}
\end{table}

\end{document}